\documentclass{article}
\usepackage{LaThuileFPSpro}
\begin{document}
\title{ 
ANTARES: TOWARDS A LARGE UNDERWATER NEUTRINO EXPERIMENT  }
\author{
  M. Spurio, on behalf of the ANTARES collaboration         \\
  {\em Dipartimento di Fisica and INFN- Bologna. spurio@bo.infn.it } \\
  }
\maketitle

\baselineskip=11.6pt

\begin{abstract}
After a long R\&D phase to validate its detector concept, the ANTARES (Astronomy with a Neutrino Telescope and Abyss environmental RESearch) collaboration\cite{coll} is operating the largest neutrino telescope in the Northern hemisphere, which is close to completion.
It is located in the Mediterranean Sea, offshore from Toulon in France at a
depth of $\sim$ 2500 m of water which provide a shield from cosmic rays. The detector design is based on the reconstruction of events produced by neutrino interactions. 
The expected angular resolution for high energy $\nu_\mu$ (E$>$10 TeV) is
less than 0.3$^o$. To achieve this good angular resolution, severe requirements on the time resolution of the detected photons and on the determination of the relative position of the detection devices must be reached.

The full 12-line detector is planned to be fully operational during this year. At present (April 2008) there are 10 lines taking data plus an instrumented line deployed at the edge of the detector to monitor environmental sea parameters.
This paper describes the design of the detector as well as some results obtained during the 2007 5-line run (from March to December).
\end{abstract}
\newpage
\section{Scientific motivation}
The main purpose of the ANTARES experiment is the detection of high energy neutrinos from galactic or extragalactic sources. Neutrinos are neutral and weak
interacting particles and they can escape from astrophysical sources, delivering direct information about the processes taking place in the core of cosmic objects.
The main scenario for the astrophysical production of high energy neutrinos involves the decay of charged pions in the beam dump of energetic protons in dense matter or photons field\cite{bere}. 
A deep connection exists between charged cosmic rays, high energy $\gamma$ emission and $\nu$ production on beam dump models. Candidates for neutrino sources are in general also $\gamma$-ray sources, since most of the mechanisms that produce neutrinos also produce high-energy photons and cosmic rays. Indeed, rather stringent limits on the diffuse neutrino flux are based on this connection (see sec. 7). 

There are many candidate neutrino sources in the cosmos; among them, supernova remnants, pulsars and micro-quasars in the Galaxy. Possible extragalactic sources include active galactic nuclei\cite{halzas}  and $\gamma$-ray burst emitters\cite{piran}. For such processes the neutrino energy scale is $10^{12}$ to $10^{16}$ eV. 
Neutrino sources that cannot be individually resolved or neutrinos 
produced in the interactions of cosmic rays with intergalactic matter or 
radiation produce a diffuse neutrino flux. This can be studied for neutrino energies in excess of $10^{14}$ eV.

ANTARES is also suited for the search of dark matter in the form of WIMPs (Weakly Interacting Massive Particles). As an example in the case of supersymmetric theories with R-parity 
conservation, relic neutralinos are predicted to concentrate in the 
centre of massive bodies such as the Earth, the Sun or the Galaxy. At these sites neutralino annihilations and the subsequent decays of the resulting particles may yield $\nu$ with energies up to $10^{10} \div 10^{12}$.

This paper describes the design of the ANTARES detector, as well as the experience and results obtained from the 5-line run (March to December 2007).

\section{ The ANTARES projet}

The ANTARES project \cite{coll} started in 1996. Today it involves about 180 physicists, engineers and sea-science experts from 24 institutes of 7 European countries. The experiment is based on the reconstruction of the direction and energy of neutrinos by detecting the Cherenkov light from particles produced in neutrino interactions. Since the neutrino interaction probability is extremely low, a huge detection volume is required to have a reasonable number of events. Secondary charged particles from cosmic rays represent the main physical background. In order to reduce it by several orders of magnitude, a large shield of kilometres of water is required.

\begin{figure}[t]
  \vspace{7.0cm}
  \includegraphics{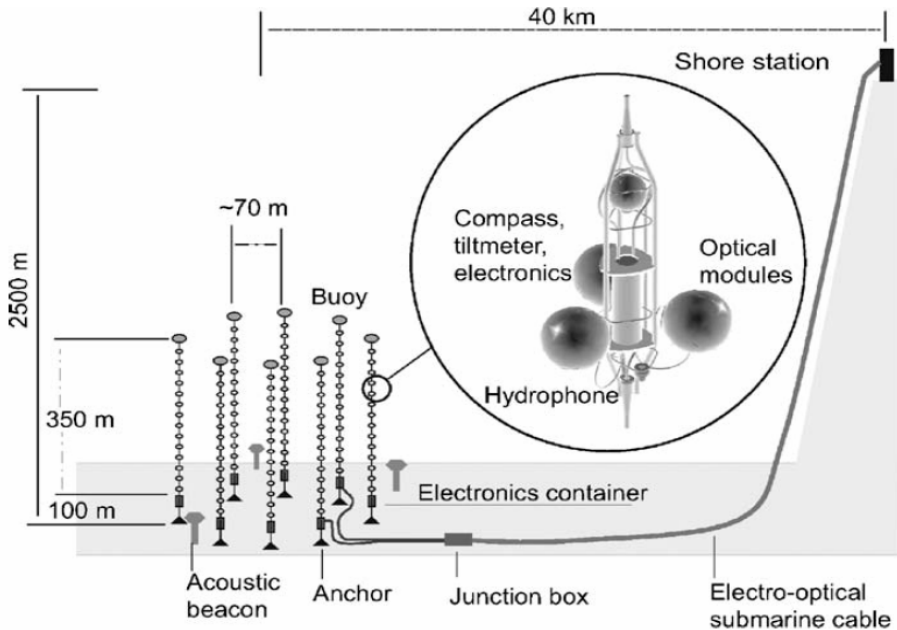}
  \caption{\it
    Schematic view of the ANTARES detector
    \label{detector} }
\end{figure}

From 1996 to 1999 an extensive R\&D program has been successfully performed to prove the feasibility of the detector concept\cite{anta1}. Site properties that have been extensively studied are: the optical properties of the surrounding water; the biofouling on optical surfaces; the optical backgrounds due to bioluminescence and to the decay of the radioactive salts present in sea water; the geological characteristics of its ground. These studies have lead to the selection of the present site, 40 km off La Seyne-sur-Mer (France) at 2475 m depth.

The full detector, which is almost completed, will consists of 12 lines made of mechanically resistant electro-optical cables anchored at the sea bed at distances of about 70 m one from each other, and tensioned by buoys at the top. Figure \ref{detector} shows a schematic view of the detector array indicating the principal components of the detector.
Each line has 25 storeys, and each storey (inset in figure) contains three optical modules (OM) and a local control module for the corresponding electronics. The OM are arranged with the axis of the PMT 45$^o$ below the horizontal.  In the lower hemisphere there is an overlap in angular acceptance between modules, permitting an event trigger based on coincidences from this overlap.

On each line, and on a dedicated instrumented line, there are different kinds of sensors and instrumentation (LED beacons, hydrophones, compasses/tiltmeters) for the timing and position calibration. The first storey is about 100 m above the sea floor and the distance between adjacent storeys is 14.5 m.
The instrumented volume corresponds to about 0.04 km$^3$. 

\begin{figure}[t]
  \vspace{6.0cm}
  \includegraphics{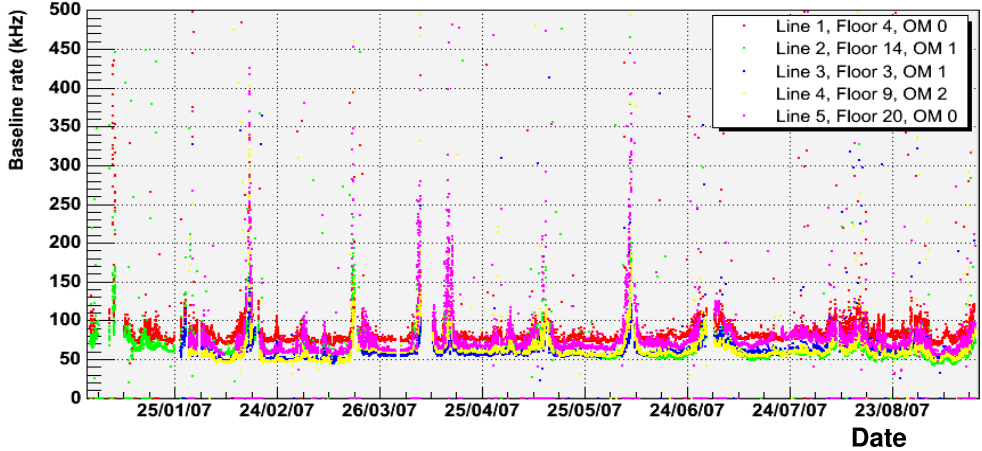}
  \caption{\it
    Counting rates of optical modules (in kHz) from January to September, 2007. The five coloured points show the rate on 5 different storeys (numbers 4, 14, 3, 9 and 20) of the five lines. Storey 25 is the uppermost. The holes represent periods of calibration or interruptions in the data taking. 
    \label{baseline} }
\end{figure}

The basic unit of the detector is the optical module (OM), consisting of a photomultiplier tube, various sensors and the associated electronics, housed in a pressure-resistant glass sphere \cite{antaom}. Its main component is a 10-inch hemispherical photomultiplier model R7081-20 from Hamamatsu (PMT) glued in the glass sphere with optical gel. A $\mu$-metal cage is used to
shield the PMT against the Earth magnetic field. Electronics inside the OM are the PMT high voltage power supply and a LED system used for internal calibration.

At present (April 2008) there are 10 lines taking data (plus the instrumented line, IL). The two remaining lines will be deployed and connected during 2008. The total sky coverage is 3.5$\pi$ sr, with an instantaneous overlap of 0.5$\pi$~sr with that of the IceCube experiment. The Galactic Centre will be observed 67\% of the day time. 

\section{ The Data Acquisition system}

The Data acquisition (DAQ) system of ANTARES is extensively described in\cite{antadaq}. 
The PMT signal is processed by an ASIC card (the Analogue Ring Sampler, ARS) which measures the arrival time and charge of the pulse. 
On each OM, the counting rates exhibit a baseline dominated by optical background due to sea-water $^{40}$K and bioluminescence coming from bacteria as well as bursts of a few seconds duration, probably produced by
bioluminescent emission of macro-organisms. Figure \ref{baseline} shows the counting rates recorded by five OMs located on different storeys of each of the 5 lines during the 2007 run. The average counting rate increases 
from the bottom to the upper layers. The baseline is normally between 50 to 80 kHz. There can be large variations of the rate, reaching hundreds of kHz for some small periods.
\begin{figure}[t]
  \vspace{6.0cm}
  \includegraphics{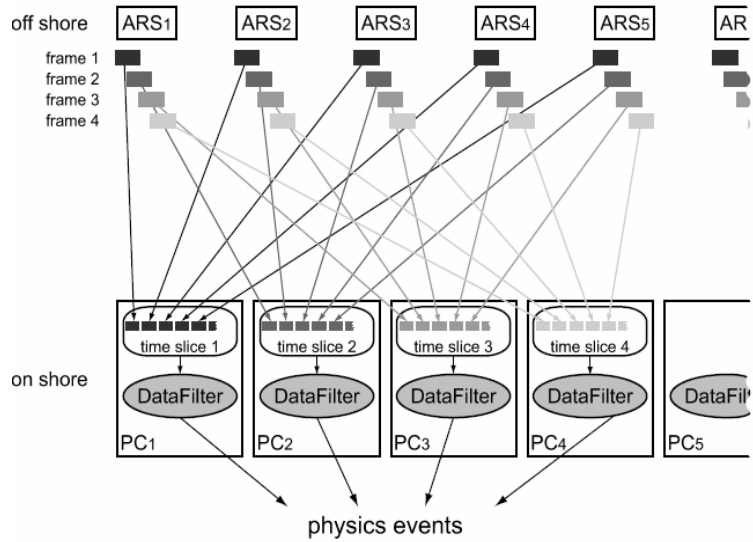}
  \caption{\it
    Scheme of the data processing based on time slices. All frames belonging to the same time window are sent to a single PC and form a time slice. The DataFilter
program running on each PC processes the data in the time slice. All physics events are stored on disk.
    \label{daq} }
\end{figure}

The optical modules deliver their data in real time and can be remotely controlled through a Gb Ethernet network. Every storey is equipped with a Local Control Module (LCM) which contains the electronic boards for the OM signal processing, the instrument readout, the acoustic positioning, the power system and the data
transmission. Every five storeys the Master Local Control Module (MLCM) also contains an Ethernet switch board which multiplexes the DAQ channels from the other storeys. 
At the bottom of each line, the Bottom String Socket (BSS) is equipped with a String Control Module (SCM) which contains the local readout and DAQ electronics, as well as the power system for the whole line. Both MCLM and SCM include a Dense Wavelength Division Multiplexing (DWDM) system used for data transmission in order to merge several 1Gb/s Ethernet channels on the same pair of optical fibres using different laser wavelengths.
The lines are linked to the junction box by electro-optical cables which are connected using a unmanned submarine. A standard deep sea telecommunication cable links the junction box with the shore station where the data are filtered and recorded.

The trigger logic in the sea is planned to be as simple and flexible as possible. All OMs are continuously read out and the digitized information ($hits$) sent to shore. On-shore, a dedicated computer farm performs a global selection of $hits$ looking for interesting physics events (DataFilter). This on-shore handling of all raw data is the main challenge of the ANTARES DAQ system, because of the high background rates. 

A $hit$ is a digitized PMT signal above the ARS threshold, set around 1/3 of the single photoelectron level (Level 0 hits, L0). The data output rate is from 0.3 GB/s to 1 GB/s, depending on background and on the number of active strings.  A subset of L0 fulfilling particular conditions were defined for triggering purposes (Level 1 hits, L1). This subset corresponds either to coincidences within 20ns of L0 hits on the same triplet of OM of a storey, or a single high amplitude L0 (typically $>$ 3 p.e.). 
The DataFilter processes all data online and looks for a physics event by searching a set of correlated L1 hits on the full detector on a $\sim 4 \ \mu s$ window. When an event is found, all L0 hits of the full detector during the time window are written on disk, otherwise the hits are thrown away.
Each DataFilter program running on a PC, see Figure \ref{daq}, has to 
finish processing a $\sim 100$ms time slice before it receives the next. This imposes an optimisation of the DataFilter programs 
in terms of processing speed and determines the specifications and number of the 
PCs required for online data processing. 

During the 5-line data taking period, the trigger rate was a few Hz. The rate of reconstructed atmospheric muons is around 1 Hz. When ANTARES receives an external GRB alerts\cite{antadaq}, all the activity of the detector is recorded for a few minutes. In addition, untriggered data runs were collected on a weekly basis. This untriggered data subset is used
to monitor the relative PMT efficiencies, as well as to check the timing within a storey, using the $^{40}$K activity.
The coincidence rate of the Cherenkov photons coming from a single $^{40}$K decay  on 2 PMTs of a storey is estimated by a Monte Carlo calculation which include the simulation of the OM, to be ($13\pm4$) Hz. This is in very good agreement with the measured value of $(14.5 \pm 0.4)$ Hz. 

Contrary to the $^{40}$K background, the bioluminescence suffers from seasonal and annual variations, see Figure \ref{baseline}. Since September 2006 the mean rate is below 100 kHz 75\% of the time. A safeguard against bioluminescence burst is applied online by means of a high rate veto, most often set to 250 kHz.

\section{The time and positioning calibration systems}

The reconstruction of the muon trajectory is based on the differences of the arrival times of the photons between optical modules (OMs). ANTARES is expected to achieve very good angular resolution ($< 0.3^o$ for muon events above 10 TeV). The pointing accuracy of the detector is determined largely by the overall timing accuracy of each event. It is necessary to monitor the position of each OM with a precision of $\sim$ 10 cm (light travels 22 cm per ns in water). The  pointing accuracy thus is limited by: $i)$ the precision with which the spatial positioning and orientation of the OM is known; $ii)$ the accuracy with which the arrival time of photons at the OM is measured; $iii)$ the precision with which local timing of individual OM signals can be synchronised with
respect to each other. 

The lines are flexible and move with the sea current, with displacements being a few metres at the top for a typical sea current of 5 cm/s. The positions of the OMs are measured in real-time, typically once every few minutes, with a system of acoustic transponders and receivers on the lines and on the sea bed together with tilt meters and compasses. The shape of each string is reconstructed by performing a global fit based on all these information. Additional information needed for the line shape reconstruction are the water current flow and the sound velocity in sea water, which are measured using different equipments: an Acoustic Doppler Current Profiler; a Conductivity-Temperature-Depth sensors; a Sound Velocimeter. 

The time resolution between OMs is limited by the transit time spread of the signal in the PMTs (about 1.3 ns) and by the scattering and chromatic dispersion of light in sea water (about 1.5 ns for a light propagation of 40 m).
The electronics of the ANTARES detector is designed to contribute less than 0.5 ns to the overall time resolution. 

Complementary time calibration systems are implemented to measure and monitor the relative times between different components of the detector  at the one ns level. These time calibrations are performed by: 

$i)$ the internal clock calibration system. It
consists of a 20 MHz clock generator on shore, a clock distribution system and a clock signal transceiver board placed in each LCM. The system also includes the
synchronisation with respect to Universal Time, by assigning the the GPS timestamp to the data. This 
system provides the absolute timing up to the level of each LCM.

$ii)$  The internal Optical Module LEDs: inside each OM there is a blue LED attached to the back of the PMT. These LEDs are used to measure the relative variation of the PMT transit time using data from dedicated runs. 

$iii)$  The Optical Beacons \cite{antabeacon}, which allow the relative time calibration of different OMs by means of independent and well controlled pulsed light sources distributed throughout the detector. 

$iv)$ Several thousands of down-going muon tracks are detected per day. The hit time residuals of the reconstructed muon tracks can be used to monitor the
time offsets of the OM,  enabling an overall space-time alignment and calibration cross-checks.

\section{Atmospheric muons }
Although ANTARES is located under a large water depth, a great number of atmospheric muons reach the active volume. They are produced in the decay of charged mesons produced at 10-20 km height by the interactions of primary cosmic rays (CR) with atmospheric nuclei. They represent the most abundant signal in any neutrino telescope and can be used to calibrate the detector and to check the simulated Monte Carlo response to the passage of charged particles. On the other side, atmospheric muons constitute the major background source, mainly because they can incorrectly be reconstructed as upward-going particles mimicking high energy neutrino interactions. Muons in bundles seem to be particularly dangerous.

\subsection{Monte Carlo simulations}

In ANTARES, two different Monte Carlo simulations are used to generate atmospheric muons: one based on a full Corsika simulation, and another based on a parameterisation of the underwater muon flux.

{\bf Full Monte Carlo.} The  full Monte Carlo simulation\cite{brunner} is based on Corsika\cite{corsika}, starting from the interactions of primary CR with atmospheric nuclei. An angular range from 0 to 85 degrees and an energy per nucleon range from 1~TeV to 100 PeV are considered
for the primaries. At lower energies the produced muons cannot reach anymore the detector whereas at higher energies the primary flux becomes negligibly small. The package QGSJET\cite{qgsjet} for the hadronic shower development has been chosen, because it has the lowest CPU need among several packages with equivalent results. As output of the first step of the simulation, muon events at the sea surface are obtained. 
In the next step the muons are propagated to the detector using the muon propagation program MUSIC\cite{music}, which includes all relevant muon energy loss processes.
At the end of this second step muons on the surface of a virtual underwater cylinder ({\it can}) are obtained. The {\it can} defines the limit inside which charged particles are propagated using GEANT-based programs, producing also Cherenkov photons\cite{brunner}. Then, the background (extracted from real data) is added and the events are feed to a program which reproduces the DataFilter trigger logic. After this step, the simulated data have the same format as the real ones.
\begin{figure}[t]
  \vspace{6.0cm}
  \includegraphics{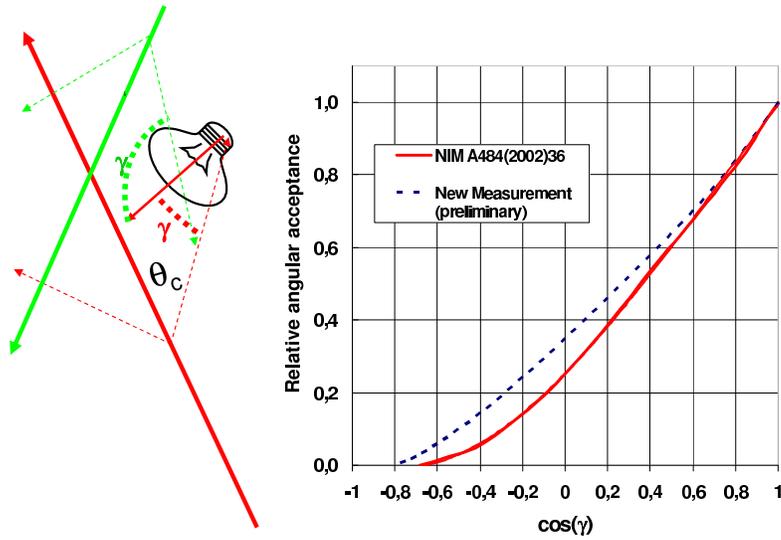}
  \caption{\it
    Left: sketch of the photon detection. The Cherenkov angle in water is $\sim 42^o$: upward (downward) going muons produce mostly photons arriving with an angle $\gamma$ smaller (larger) than $90^o$ with respect to the PMT axis. Right: measured acceptance of the ANTARES OM. A recent measurement shows a larger acceptance with respect to that with an older configuration.
    \label{om} } \end{figure}

The  main advantage of the full Monte Carlo simulation is that it is done
with a simple $E^{-\gamma}$  spectrum for the primary flux for all nuclei.
This allows a later re-weighting with any chosen primary flux model. The drawback is that a very large amount of CPU time is needed.

{\bf Monte Carlo with parametric formulas.} A second data set is generated using parametric formulas\cite{mupage0}, which allow a fast generation of a very  large sample of atmospheric muons. 

The used parameterisation of the flux of underwater muon bundles is based on a full Monte Carlo simulation of CR interaction and shower propagation in the atmosphere using the HEMAS code\cite{hemas}, with DPMJET\cite{dpmjet} to calculate the hadronic shower development.
The adopted primary CR flux is an un-published model which reproduces the muon flux (single and multiple muons) and energy spectrum as measured by the MACRO experiment\cite{macro}. The muons reaching the sea level are then propagated using MUSIC down to $5.0\ km \:w.e.$ The characteristics of underwater muon events (flux, multiplicity, radial distance from the axis bundle, energy spectrum) are described with multi-parameter formulas in the range $1.5\div 5.0\ km \:w.e.$ and up to $85^\circ $ for the zenith angle. In particular, the energy spectrum of muons depends on the vertical depth $h$, on the zenith angle $\theta$, on the muon multiplicity in the shower $m$ and on the distance of the muon from the shower axis $R$.

Using this parameterization, an event generator (called MUPAGE) was developed\cite{mupage1} in the framework of the KM3NeT project\cite{km3net} to generate underwater atmospheric muon bundles. In the case of ANTARES, the events are generated on the {\it can} surface. Then, the muons are propagated with the production of Cherenkov light, the background added and the events fed to the trigger logic as in the case of the full Monte Carlo. 

The main advantage of this simulation is that a large sample is produced with a relatively small amount of CPU time (much less than the time needed to simulate the Cherenkov light inside the {\it can}). A data set with a livetime equivalent to one month, which is used to compare data and MC, required 300 hours of CPU time on a 2xIntel Xeon Quad core, 2.33 GHz processor. The drawback is that the primary CR composition is fixed, and the events cannot be re-weighted. 

A larger data set of more energetic atmospheric muons, equivalent to one year of livetime, is generated to study the background rejection criteria in the search of diffuse flux of high energy neutrino (E$_\nu >$ 100 TeV). This sample required 232 hours of CPU time (with the aforementioned processor), when a cut on the total energy of the underwater muons is applied ($E_{total} > 3$ TeV).

\subsection{Results for the 5-line run}

As expected, atmospheric muons were an important tool to monitor the detector status and to check the reliability of the simulation tools and data taking. 

The early comparison of atmospheric muons shows a large discrepancy between data and Monte Carlo, the MC rate being about 1/3 of that measured. This pushed us for systematic checks of all sections in our Monte Carlo simulations (water absorption length and scattering models; Cherenkov light production; tracking algorithm procedure; description of the optical module effective area, etc.) as well as in the analysis data chain (efficiency of the trigger algorithm, etc.). 

The main problem was found in the description of the optical module response. 
The three PMTs in each storey are oriented with axis 45$^o$ below the horizontal. They detect light with high efficiency from the lower hemisphere (from where neutrinos are expected), and has some acceptance for muons coming from above the horizontal. The OM angular acceptance used in the MC code (red line in Figure \ref{om}) was measured\cite{antaom} with an old configuration. It is broad, falling to half maximum at ±70$^o$ from the axis of the PMT. When re-measured with the present OM configuration (blue line in Figure \ref{om}), it shows higher values for incoming photons angles, $\gamma$, larger than $\sim 60^o$. As a consequence, the number of MC reconstructed atmospheric muons increases by 200-300\%, while upward going particles (neutrino induced) increases at most by $\sim$ 20\%. Another small inefficiency was found in the data acquisition by comparing the distribution of the number of triggering hits in data and MC.  

\begin{figure}[t]
  \vspace{7.0cm}
  \includegraphics{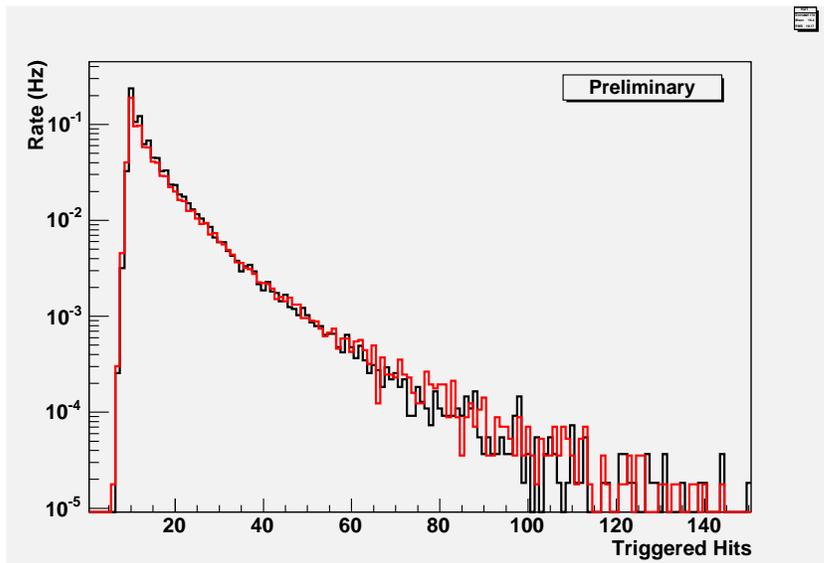}
  \caption{\it
The trigger process (DataFilter) act on a subset of hits, the so called L1 hits (see text). When a sufficient number of correlated L1 hits is found, the data are considered as due to a physics event and all information from a $\sim 4 \mu s$ time window written to disk. The triggered hits are those hits which enabled the trigger logic. The figure shows the triggered event rate (in Hz) versus the number of triggered hits in the event. Black histogram: data. Red histogram: Monte Carlo (atmospheric muons). 
    \label{trighits} }
\end{figure}

\begin{figure}[t]
  \vspace{9.0cm}
  \includegraphics{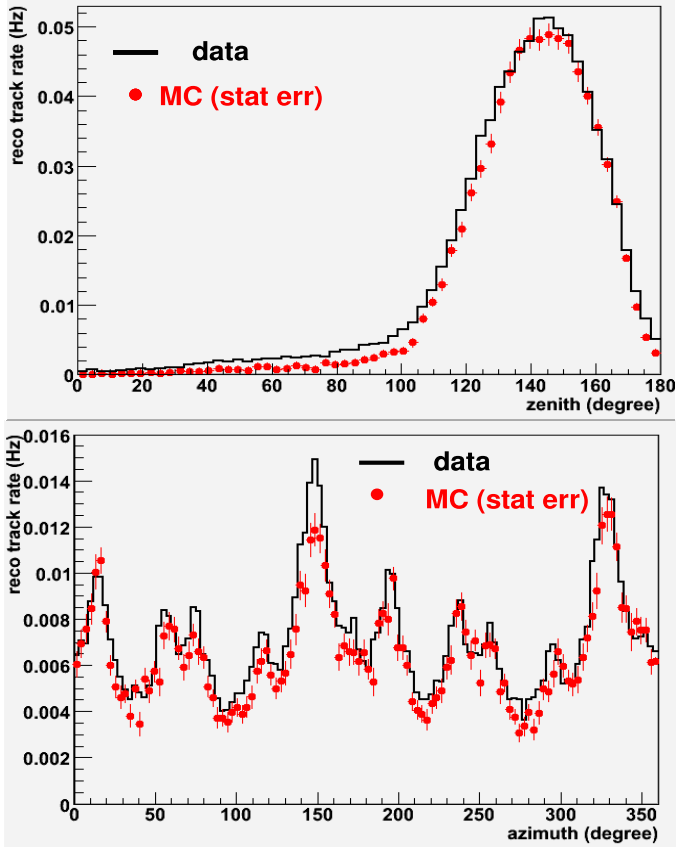}
  \caption{\it
Zenith (upper plot) and azimuth (lower plot) distribution of atmospheric muons detected in the 5-line run. Black: experimental data. Red: MC simulation. The shape of the azimuth distribution reflects the geometry of the 5 line detector. 
    \label{zenazi} }
\end{figure}

The results after the checks in the Monte Carlo and in the data acquisition chain are shown in Figure \ref{trighits} and \ref{zenazi}.
Figure \ref{trighits} shows the distribution of the number of hits used by the  trigger process (DataFilter) to trigger the event in data and Monte Carlo (atmospheric muons from MUPAGE\cite{mupage1}). 
Figure \ref{zenazi} shows the zenith and azimuth angle distribution of reconstructed events (without any quality cuts) in data (black) and MC (red). 
At present, a systematic error (constant for all bins) of $\sim 40\%$ which include the uncertainties on the interaction model but not that on the primary CR composition is estimated for the MC predictions. In the figure, the data and MC are not normalized but the agreement in the integrated data should be consider fortuitous, due to the used CR composition model.

The quality cuts in the reconstruction applied for the selection of 
neutrino candidates are needed in order to reduce the badly 
reconstructed downward going events mimicking upward going tracks (see 
next section).

\section{Neutrino candidates from the 5 Line data set}

The bulk of triggered events are due to downward going atmospheric muon 
A first, very preliminary, analysis has been performed on some high quality data, ignoring all  storeys positioning aspects and assuming straight lines. The alignment, results of line shape fits using slow control positioning data available every 6 minutes, is currently implemented in track reconstruction, and will soon allows for a much efficient determination of the neutrino candidates. 

\begin{figure}[t]
  \vspace{6.0cm}
  \includegraphics{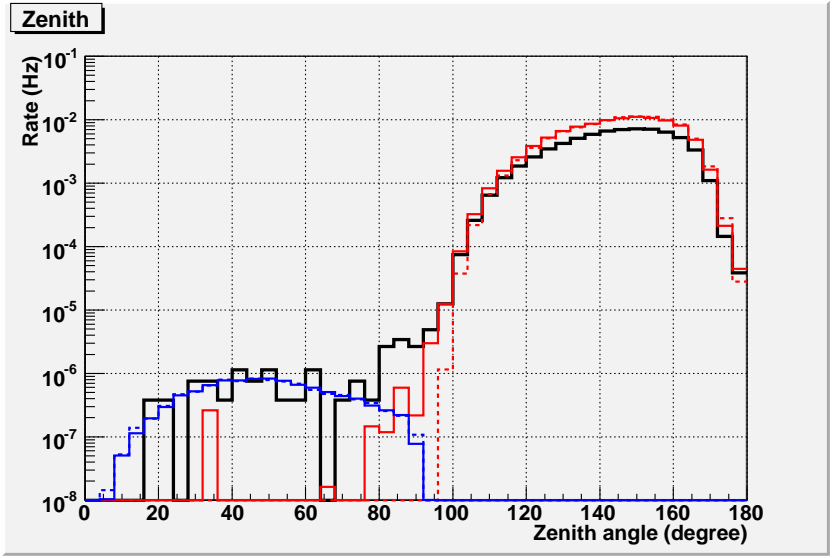}
  \caption{\it
    Zenith angle distribution of detected events after quality cuts. Upgoing events have zenith angle$<90^o$. Black lines represent data. Red stands for atmospheric muons, and blue indicates atmospheric neutrinos. MC-truth is shown as dotted lines, full lines are reconstruction results. For this analysis, the dynamic positioning of each OM was NOT used. 
    \label{zenith} }
\end{figure}

Figure \ref{zenith} shows the comparison between reconstructed data and Monte Carlo. The data sample consists of 36.8 days of active time from selected runs between 01/02/2007 and 25/05/2007. The atmospheric muons are from the Corsika-based Monte Carlo, with the primary CR flux of \cite{horandel}. The neutrino events are simulated using the Bartol flux\cite{bartol}. Only events detected at least by two lines and with at least 6 floors are considered. The integrated rates (after quality cuts) as shown on the plots are 0.07 Hz for data and 0.10 Hz for atmospheric muons. Restricting to the upward going hemisphere (neutrino candidates) this becomes 1.4 per day for data, 0.11 per day for atmospheric muons and 0.84 per day for atmospheric neutrinos. When this information is included, the number of reconstructed events per day is expected to increases by at least a factor of three, with an higher angular resolution. 

\section{Expected performances}

The expected performances of the full 12-line detector have been estimated by
computer simulation. The capabilities of the telescope can be characterized by several quantities. For instance, the muon effective area gives the ratio of the 
number of well-reconstructed ({\it selected}) muon events to the incoming muon flux. 
The effective area increases with energy: it is 0.02(0.04) km$^2$ for $E_\nu = 10(100)\ TeV$ and reaches 0.08 km$^2$ for neutrino energies larger than $10^4$ TeV. 
These values assume {\it selected} events, in such a way that the median of the distribution of the angular difference in space between the reconstructed muon track and the original parent neutrino is better than $0.3^o$.
The angle between the parent neutrino and the muon is dominated by kinematics effects up to around 10 TeV. Above that energy, the instrumental resolution is the limiting factor. 
A good angular resolution helps to reject the background whenever the 
source position is relevant, as is the case in the search of point-like 
sources. 

\begin{figure}[t]
  \vspace{8.0cm}
  \includegraphics{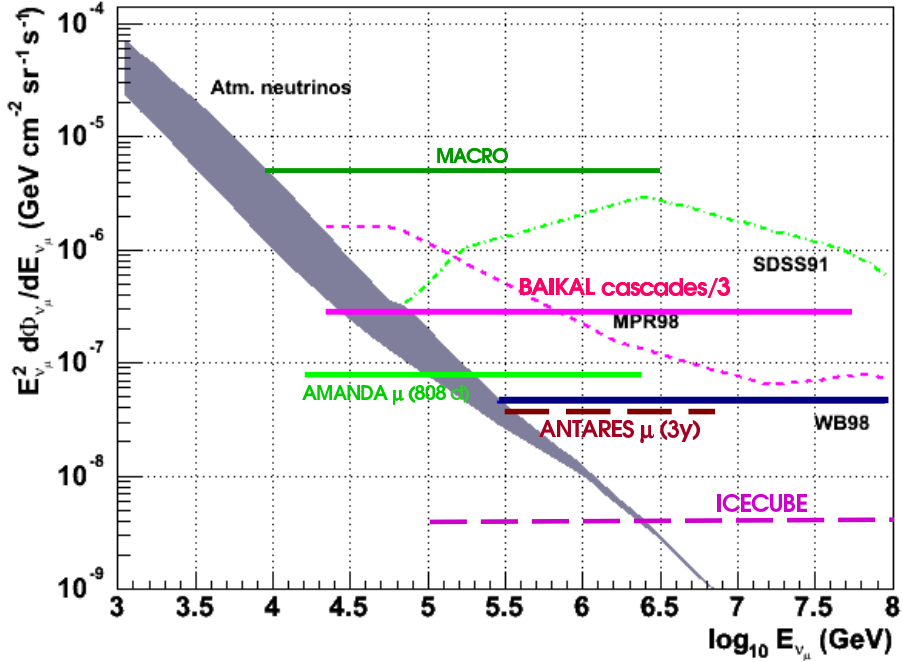}
  \caption{\it Diffuse flux scaled to an $E^{-2}$ spectrum as a function of the neutrino energy. The upper limit that ANTARES can set in 3 year is indicated together with the expected atmospheric flux, some theoretical predictions and limits from other experiment.  
    \label{diffuse} }
\end{figure}

The energy of the crossing muon or of secondary particles generated by neutrino interactions inside the instrumented volume is estimated from the amount of light deposited in the PMTs. Several estimators based on different techniques were developed\cite{romeyer}. MC studies show that this resolution  is between $log_{10}(\sigma_E/E)=0.2\div 0.3$ for muons with energy above 1 TeV. 
The event energy measurement is a mandatory requirement for the study of the diffuse flux of high energy neutrinos. 
The link between the extra-galactic sources of cosmic rays, gamma-rays and neutrinos leads to severe limits on the neutrino diffuse flux expressed in the Waxman and Bahcall (WB98) upper limit $E^2\Phi<4.5\times 10^{-8}\ GeV\
cm^{-2} s^{-1} sr^{-1}$ \cite{wb}. Monte Carlo simulations indicate that after 3 year of data taking ANTARES can set an upper limit for diffuse fluxes of $E^2\Phi<3.9\times 10^{-8}\ GeV\ cm^{-2} s^{-1} sr^{-1}$, just below the WB98 upper limit (Figure~\ref{diffuse}). This value marks a limit for a list of known candidate sources, but must be corrected to take into account neutrino oscillations.

The ANTARES sensitivity to point-like sources is estimated as a function of the declination. The 90\% C.L. upper limit for the $\nu_\mu+\overline{\nu_\mu}$ flux from point-like sources we can set in case of a null signal after one year of data taking is $E^2dN/dE_\nu =4\times 10^{-8} GeV cm^{-2}s^{-1}$ for a declination of $\delta=-90^o$ and rises to $1.5\times 10^{-7} GeV cm^{-2}s^{-1}$ for $\delta=+40^o$. These limits improve those of SuperKamiokande and MACRO from the Southern sky and are comparable to those obtained by AMANDA II in 1001 days from the Northern sky.

\section{Prospective and conclusions}

ANTARES is at present the largest neutrino observatory in the Northern hemisphere, which represents a privileged sight of the most interesting areas of the sky like the Galactic Centre, where neutrino source candidates are expected. 
It is able to explore the Southern sky hemisphere in the search for astrophysical neutrinos with a sensitivity much better than any other previous experiments. 

ANTARES is also the most advanced pilot neutrino telescope in the Mediterranean sea toward the km3-scale telescope, with a strong relationship and cooperation with the NEMO\cite{migneco} project. 
Most of the theoretical models put the sensitivity for discovering neutrino sources at a level for which a telescope $\sim$ 50 times larger than ANTARES is required (or 3 times IceCube, which observes the complementary sky region).
While ANTARES is taking and analyzing data, some of the collaboration activities are continuing in the framework of the KM3NeT project\cite{km3net}. 

The KM3NeT Design Study is a 3-year project (started in 2006) which is founded by the EU within the VI Framework Programme. The Design Study objective is to produce a Technical Design Report by the Summer 2009. For this report, decisions on the implementations of the different components of the Neutrino Telescope must be taken, with a full costing of the solutions.  A Conceptual Design Report will be released at the end of this month.  
In parallel, in March 2008, the Preparatory Phase of the KM3NeT project has started and will continue until March 2011. In this phase, a small-scale engineering model of the detection unit and the sea-floor infrastructure will be produced. The final selection of the site will be pursued in the framework of the Preparatory Phase and will likely involve decisions at the political level.

\end{document}